\documentclass[aip,apl,reprint]{revtex4-1}

\usepackage{amsmath,amssymb}
\usepackage{graphicx}
\usepackage{dcolumn}
\usepackage{bm}

\usepackage{color}

\newcommand{\beq}[1]{  \begin{equation} \label{#1} }  
\newcommand{\eeq}{     \end{equation}}  
\newcommand{\bal}[1]{\begin{align} \label{#1} }

\def\div{\operatorname{div}}

\begin{document} 

\title{Comment on ``Design of acoustic devices with isotropic material via conformal
	transformation" [Appl. Phys. Lett. 97, 044101 (2010)]}  

\author{ Andrew N. Norris} 
\affiliation{Mechanical and Aerospace Engineering, Rutgers University, Piscataway NJ 08854}

\date{\today}

\begin{abstract}

The paper presents incorrect formulas for the  density and bulk modulus under a  conformal transformation of coordinates.    The fault lies with an  improper assumption of constant acoustic impedance.

\end{abstract}


\maketitle



The wave equation for an acoustic medium with uniform density  $\rho_0$ and bulk modulus $\lambda_0$ is, 
assuming time dependence $e^{-i\omega t}$, 
\beq{1}
\nabla\cdot \nabla p + \omega^2 (\rho_0/\lambda_0)\, p = 0.
\eeq 
Ren et al.  \cite{Ren10} demonstrated that under the  conformal transformation $x^i  \rightarrow x^{i'}$, $z (= x^1+ix^2)  \rightarrow \zeta (z)$, eq.\ \eqref{1} becomes, for $p' ( x^{i'}) = p (x^i)$, 
\beq{3}
\nabla '\cdot \nabla ' p'+ \omega^2 (\rho '/\lambda ')\,  p^\prime = 0 , 
\eeq 
where 
\beq{4}
\lambda ' / \rho '  =  |   \zeta ' (z)|^2\, \lambda_0/\rho_0 .
\eeq 
The authors  \cite{Ren10}  then imposed the constraint that the impedance in the transformed domain remains unchanged, i.e. 
\beq{5}
\lambda '   \rho '  =  \lambda_0 \rho_0, 
\eeq 
which together with \eqref{4} implies the identities (eq.\ (6) of \cite{Ren10})
\beq{6}
  \rho '  = \rho_0 / |   \zeta ' (z)|, \quad
	\lambda '  =  \lambda_0  |   \zeta ' (z)| .
\eeq
The expressions \eqref{6} are not correct as we explain next.   

The  equations of motion for  an acoustic fluid of density 
$\rho '$ and bulk modulus $\lambda '$ are
\beq{7}
-\omega^2 \rho ' \mathbf{u}' = -\nabla ' p',
\quad
 p' = - \lambda '\nabla '\cdot \mathbf{u}'.
\eeq
Eliminating the displacement $\mathbf{u}'$ yields
\beq{8}
\rho ' \nabla '\cdot (1/\rho ')\nabla ' p'+ \omega^2 (\rho '/\lambda ')\,  p^\prime = 0 . 
\eeq 
Comparison of eqs.\ \eqref{3} and \eqref{8} implies that the former  is the reduced equation for the pressure in an acoustic fluid only if $\rho '$ is constant.   That is, 
eqs.\ \eqref{3} and \eqref{4} represent the equation of motion of an acoustic medium with 
 \beq{9}
  \rho '  = c \rho_0  , \quad 
	\lambda '  =  c \lambda_0  |   \zeta ' (z)|^2 ,
\eeq
for constant $c>0$.  These are the correct expressions for the bulk modulus and density in the transformed medium.

The general theory of transformation acoustics allows for the possibility of non-unique expressions for the material parameters in the transformed domain \cite{Norris08b}.  The material descriptions range  from what is known as inertial fluids, with anisotropic density and scalar bulk modulus, to pentamode behavior with elastic constitutive response and generally anisotropic density, plus a spectrum of possibilities in between \cite{Norris09}.  In the inertial fluid limit the transformed bulk modulus and density tensor  are
(eq.\ (2.8) of Ref.\ \onlinecite{Norris08b}) 
\beq{-1}
\lambda '  =    \lambda_0 \det {\bf F}, \quad
{\pmb \rho} ' = \rho_0 (\det {\bf F}) \, ({\bf F}{\bf F}^T)^{-1}, 
\eeq
where  $F_{i' i} = \partial x^{i'}/\partial x^i$.   At the other extreme, the pentamode material is defined by a density tensor ${\pmb \rho}$ and a fourth order elasticity tensor with
components $C_{ijkl}$ (eq.\ (4.3) of Ref.\ \onlinecite{Norris08b}) 
 \beq{-2}
C_{ijkl}  =    \lambda '\, S_{ij}S_{kl}, \quad
{\pmb \rho} =   {\bf S} {\pmb \rho} ' {\bf S}, 
\eeq
where $\lambda '$, ${\pmb \rho} '$ are given by \eqref{-1} and  ${\bf S} = {\bf S}^T$ is positive definite satisfying $\div {\bf S}=0$, but otherwise arbitrary.  For the special case of  a conformal transformation we have $\det {\bf F}  = |   \zeta ' (z)|^2$,  ${\bf F}{\bf F}^T= |   \zeta ' (z)|^2{\bf I}$ where ${\bf I}$ is the identity.  Then 
eq.\ \eqref{-1} reduces to \eqref{9}, apart from the factor $c$ which could be incorporated into ${\bf F}$, and taking $ {\bf S} = {\bf I}$ eq.\ \eqref{-2} also reduces to \eqref{9} since 
$C_{ijkl} = \lambda ' \delta_{ij}\delta_{kl}$ corresponds to the acoustic fluid of bulk modulus $\lambda '$.
In conclusion,  when the transformation is  conformal the parameters are isotropic, with both the inertial  and the pentamodal models  yielding   $  \rho '$ and $\lambda '$ given by eq.\ \eqref{9}.

\begin{acknowledgments}
Thanks to N. Gokhale and A. J. Nagy for comments.
\end{acknowledgments}


%

\end{document}